\documentclass[12pt,a4paper]{article}

\RequirePackage[l2tabu, orthodox]{nag}

\usepackage{mjheppub}

\usepackage{Pmacro}

\usepackage{pst-all}

\title{\boldmath Lax matrix solution\\ of $c=1$ Conformal Field Theory}

\author{Bertrand Eynard}
\author{and Sylvain Ribault}

\affiliation{\vspace{2mm} CEA Saclay, Institut de Physique Th\'eorique}

\emailAdd{firstname.name@cea.fr}

\abstract{To a correlation function in a two-dimensional conformal field theory with the central charge $c=1$, we associate a matrix differential equation \(\Psi' = L \Psi\), where the Lax matrix $L$ is a matrix square root of the energy-momentum tensor. Then local conformal symmetry implies that the differential equation is isomonodromic. This provides a justification for the recently observed relation between four-point conformal blocks and solutions of the Painlev\'e VI equation. This also provides a direct way to compute the three-point function of Runkel-Watts theory -- the common $c\rightarrow 1$ limit of Minimal Models and Liouville theory.
}

\preprint{\jobname.tex}

\begin{document}

\maketitle
\flushbottom

\section{Introduction}

Two-dimensional conformal field theories have an infinite-dimensional symmetry algebra, the Virasoro algebra, and can therefore be considered as integrabe models. This in principle makes them accessible to methods such as the Bethe ansatz. These methods are usually less powerful than the methods which rely specifically on conformal symmetry, such as the conformal bootstrap method. However, it is still interesting to study conformal field theories from an integrable viewpoint, because:
\begin{itemize}
 \item The integrability of a conformal field theory can survive deformations which break conformal symmetry.
\item Integrable structures such as spectral curves can synthetically encode much information on a theory. This can be particularly useful in two-dimensional conformal field theories which have no Lagrangian formulation.
\item For some calculations, integrable methods are simpler than the conformal bootstrap method, as we will see. 
\end{itemize}
Integrable structures have been investigated in particular in the cases of minimal models \cite{blz94} and Liouville theory \cite{bt09}. In both cases, the approach was to associated an integrable model to a conformal field theory. However, the relations of Liouville theory to matrix models and to gauge theories \cite{agt09} suggest another approach: to associate 
an integrable model to each correlation function of Liouville theory, so that the variables of the correlation function, such as field positions, correspond to parameters of the model, such as coupling constants. We will follow this second approach.

This approach was already shown to provide a perturbative expansion of Liouville correlation functions around the value $c=\infty$ of the central charge \cite{cer12}. This perturbative expansion is encoded in a non-commutative spectral curve, as was demonstrated by the explicit computation of the first two terms of the three-point function. Here we will use this approach for exactly solving Liouville theory at $c=1$. Liouville theory is supposed to exist as a consistent, unitary conformal field theory for $c>1$ \cite{zz95}, although its quantum gravity interpretation is clear only for $c>25$. The $c=1$ limit of Liouville theory is another consistent, unitary conformal field theory \cite{sch03}, which was originally found by Runkel and Watts as a limit of Minimal Models \cite{rw01}, as we illustrate in the following diagram:
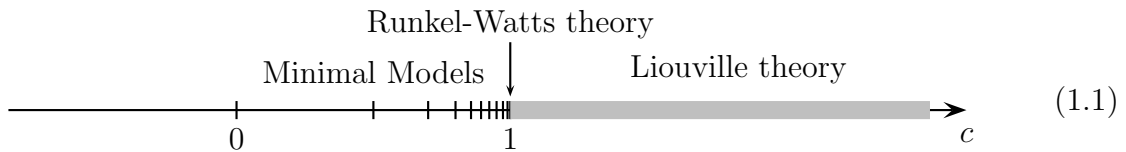
\begin{align}
 \psset{unit=.6cm}
\pspicture[shift=-1.5](-10,-1.5)(10,3)
\psline[arrowscale=2]{->}(-11,0)(10,0)
\rmultiput{\psline(0,-.2)(0,.2)}(-6,0)(-3,0)(-1.8,0)(-1.2,0)(-.86,0)(-.64,0)(-.46,0)(-.3,0)(-.16,0)(-.06,0)(0,0)
\pspolygon*[linecolor=lightgray](0,-.2)(0,.2)(9.2,.2)(9.2,-.2)
\rput[t](-6,-.4){$0$}
\rput[t](0,-.4){$1$}
\rput[t](10,-.4){$c$}
\rput[b](-3,.6){Minimal\ Models}
\rput[b](5,.6){Liouville\ theory}
\rput[b](0,1.6){Runkel-Watts\ theory}
\psline{->}(0,1.5)(0,.3)
\endpspicture
\end{align}
The reason why we focus on the $c=1$ case is the availability of powerful techniques \cite{be09} which will enable us to solve the theory with the help of a Lax matrix -- a more powerful object than the spectral curve, to which it is nevertheless directly related. 

Let us announce our plan. We begin with reviewing the Ward identities in Conformal Field Theory in Section \ref{ssecwi}, before using the ansatz of \cite{be09} for solving them in the case $c=1$ in Section \ref{ssecmat}. We use this for solving $c=1$ Conformal Field Theory, by computing four-point conformal blocks in Section \ref{secps} and the three-point correlation function in Section \ref{sectp} (which is independent from Section \ref{secps}). In particular, we rederive recent results \cite{gil12} on the relation between four-point conformal blocks and solutions of the Painlev\'e VI equation, and explain these results by noting that the ansatz of Section \ref{ssecmat}, together with local conformal symmetry of the theory, leads to an isomonodromic matrix differential equation. To compute the three-point correlation function, we compute the Lax matrix, solve the matrix differential equation, and use Seiberg-Witten equations, for which we provide a derivation from conformal symmetry in Section \ref{ssecsw}.
 In the 
concluding Section \ref{seccon} we discuss generalizations of our method.

\paragraph{Note on bibliography.} Our bibliographical references only include data which we believe are helpful to readers: authors, dates and titles of cited articles. Adding more data is made unnecessary by search engines. 

\section{The Lax matrix in \texorpdfstring{$c=1$}{c=1} conformal field theory}

Let us recall the definitions of the basic objects of two-dimensional conformal field theory, before introducing the formalism which will enable us to compute these objects.
The main observable is the $N$-point correlation function
\begin{align}
 Z_N = \la \prod_{i=1}^N V_{p_i}(z_i,\bar{z}_i)\ra\ ,
\end{align}
where $(z,\bar{z})$ are complex coordinates on the two-dimensional Euclidean plane, and $p$ is the momentum of the primary field $V_p(z,\bar{z})$. In addition to theses parameters, $Z_N$ implicitly depends on the central charge $c$ of the theory, which parametrizes the Virasoro symmetry algebra. In the conformal bootstrap approach, $Z_N$ is determined by conformal symmetry, together with axioms such as the existence of Operator Product Expansions. These axioms imply that any correlation function $Z_N$ can be decomposed into three-point functions $Z_3$, and functions called conformal blocks, which are completely determined by conformal symmetry. 
For example, the $s$-channel decomposition of the four-point function is
\begin{align}
 Z_4 = \int \frac{dp_s}{R(p_s)}\ C(p_1,p_2,p_s)C(p_s,p_3,p_4) \left|\mathcal{F}_{p_s}(p_i|z_i)\right|^2\ .
\label{zfi}
\end{align}
Here the three-point structure constant
$C(p_1,p_2,p_3)$ is the $z_i,\bar{z}_i$-independent, non-trivial factor of $Z_3$. The reflection coefficient $R(p_s)$ is also $z_i,\bar{z}_i$-independent.
The four-point, $s$-channel conformal block $\mathcal{F}_{p_s}(p_i|z_i)$ is a locally holomorphic functions of $z_i$ -- in other words, a $\bar{z}_i$-independent function. This conformal block is determined by conformal symmetry up to a $z_i$-independent normalization factor.

\subsection{Ward identities in conformal field theory \label{ssecwi}}

Conformal symmetry is conveniently encoded in the properties of a meromorphic spin-two field $T(z)$ called the energy-momentum tensor. 
Following \cite{cer12}, we use an alternative approach based on a spin-one field $J(z)$, which is locally holomorphic but can have branch cuts, and can therefore be considered multivalued as a function of $z$. From $J(z)$ it is possible not only to recover $T(z)$, but also to write Ward identities and Seiberg-Witten equations, which lead to a method for computing correlation functions. We will now review this method.
\footnote{In this Subsection, we keep the central charge $c$ arbitrary, but adopt notations which will be convenient in the case $c=1$, and are related to the notations in \cite{cer12} by 
\begin{align*}
J^{\mbox{\cite{cer12}}} = i J \quad , \quad Q^{\mbox{\cite{cer12}}} = iq \quad , \quad \alpha^{\mbox{\cite{cer12}}} = ip \quad , \quad W_m^{\mbox{\cite{cer12}}} = i^mW_m\ .
\end{align*}
}

The field $J(z)$ is defined by its self-Operator Product Expansion (OPE),
\begin{align}
 J(y)J(z) = \frac{\frac12}{(y-z)^2} + (JJ)(z) + O(y-z)\ .
\label{jj}
\end{align}
The singular part $\frac{\frac12}{(y-z)^2}$ of this OPE dictates that the modes $J_n$ defined by $J(z) = \sum_{n\in\Z} J_n z^{-n-1}$ form an  affine Lie algebra $\hat{\mathfrak{u}}_1$. The first regular term $(JJ)(z)$ encodes the definition of the field $(JJ)$ from the field $J$, as a normal-ordered product. The energy-momentum tensor can be recovered from $J$ by
\begin{align}
 T = (JJ) +q\p J\ ,
\label{tj}
\end{align}
where the parameter $q$ is called the background charge. As this relation is quadratic, $J(z)$ is double-valued as a function of $z$.
The self-OPE $T(y)T(z)$, which can be deduced from the self-OPE $J(y)J(z)$, encodes 
the Virasoro symmetry algebra with the central charge 
\begin{align}
 c = 1 - 6 q^2\ .
\label{cq}
\end{align}
The locally holomorphic field $J(z)$, together with its antiholomorphic counterpart $\bar{J}(\bar{z})$, generates the chiral algebra of conformal field theory. In theories without $\hat{\mathfrak{u}}_1$ symmetry, this chiral algebra is not a symmetry algebra, but a spectrum-generating algebra. The symmetry algebra is generated by $T(z)$ and its antiholomorphic counterpart $\bar{T}(\bar{z})$.

We now define the primary field $V_p(z,\bar{z})$ by its OPE with $J(y)$,
\begin{align}
 J(y)V_p(z,\bar{z}) = \frac{p}{y-z} V_p(z,\bar{z}) + (JV_p)(z,\bar{z}) + O(y-z)\ .
\label{jv}
\end{align}
From this definition, and the definition (\ref{tj}) of the energy-momentum tensor, it
follows that $V_p(z,\bar{z})$ is a Virasoro primary field,
\begin{align}
 T(y) V_p(z,\bar{z}) = \left(\frac{\Delta}{(y-z)^2} +\frac{1}{y-z}\pp{z}\right) V_p(z,\bar{z}) + O(1)\ ,
\label{tv}
\end{align}
with the conformal dimension
\begin{align}
 \Delta = p(p-q)\ .
\label{da}
\end{align}
The relevant physical parameter of a primary field is actually the conformal dimension, not the momentum. Two fields with the same conformal dimension must be proportional to each other. Since the conformal dimension is invariant under the reflection 
$p \mapsto q-p$ of the momentum, we must have
\begin{align}
 V_{p}(z,\bar{z}) = R(p) V_{q-p}(z,\bar{z})\ ,
\label{vrv}
\end{align}
where the number $R(p)$ is called the reflection coefficient. 
This is  apparently incompatible with the definition (\ref{jv}) of $V_p(z,\bar{z})$. Remember however that $J(y)$ is double-valued as a function of $y$: 
there should exist two determinations of $J(y)$, associated to the OPE coefficients $p$ and $q-p$, and which are exchanged by reflection.

 Let us define correlation functions involving insertions of $J(y)$,
\begin{align}
 \hat{W}_m(y_1,\cdots y_m) = \lla \prod_{i=1}^n J(y_i) \rra\ , \quad \mathrm{where} \quad \lla \mathcal{O} \rra = \frac{1}{Z_N}\la \mathcal{O} \prod_{i=1}^N V_{p_i}(z_i,\bar{z}_i)\ra\ ,
\end{align}
and let $W_m(y_1,\cdots y_m)$ be the corresponding regularized connected correlation functions, in particular
\begin{align}
 \hat{W}_1(y)&= W_1(y)\ ,
\\
\hat{W}_2(y_1,y_2) & = W_2(y_1,y_2) +\frac{\frac12}{(y_1-y_2)^2} +W_1(y_1)W_1(y_2)\ .
\end{align}
(See \cite{cer12} for the explicit definition of $W_m$ for all $m$.) 
Such correlation functions obey a system of Ward identities. These identities are the consequences for $W_m$ of the property (\ref{tv}) of the energy-momentum tensor $T(z)$, via its relation (\ref{tj}) with the field $J(z)$. The Ward identities therefore express the local conformal symmetry of the theory, and they read \cite{cer12} 
\begin{multline}
 -q\pp{y} W_{m+1}(I,y) + W_{m+2}(I,y,y)+ \sum_{J\subset I}W_{|J|+1}(J,y)W_{m-|J|+1}(I-J,y) 
\\ 
+\sum_{j=1}^m\pp{y_j}\frac{W_m(I-\{y_j\},y)-W_m(I)}{y-y_j} = -P_{m+1}(I,y)\ ,
\label{qpw}
\end{multline}
where $m\geq 0$ is an integer, $I=(y_1,y_2,\cdots y_m)$, and we define 
\begin{align}
 P_{m+1}(I,y) =  -\delta_{m,0}t(y) - D_yW_m(I) \ .
\label{pm}
\end{align}
This involves the notations 
\begin{align}
 D_y = \sum_{i=1}^N \frac{1}{y-z_i}\pp{z_i} \ ,
\label{dy}
\end{align}
and
\begin{align}
 t(y) = \lla T(y) \rra =\sum_{i=1}^N\left(\frac{\Delta_i}{(y-z_i)^2} +\frac{\beta_i}{y-z_i}\right)\ , \quad \mathrm{where} \quad \beta_i=\pp{z_i}\log Z_N\ .
\label{ty}
\end{align}
The coefficients $\beta_i$ are interpreted as accessory parameters in the limit $c\rightarrow \infty$ \cite{cer12}, and we will still call them accessory parameters for finite values of $c$. The accessory parameters are constrained by the conformal symmetry condition
\begin{align}
 t(y)\underset{y\rightarrow \infty} = O\left(\frac{1}{y^4}\right) \ ,
 \label{toy}
\end{align}
which via eq. (\ref{ty}) amounts to global Ward identities for $Z_N$. 

Finding $W_m$ by solving the Ward identities may seem difficult for several reasons:
\begin{enumerate}
\item the $m=0$ identity involves the accessory parameters $\beta_i$, which are a priori unknown,
\item the identities are differential equations in the variable $y$, so that integration constants may appear,
\item the identities form a system of infinitely many equations for infinitely many unknowns.
\end{enumerate}
These difficulties were overcome in the computation of the Liouville three-point function $Z_3$ as a perturbative series in $q^2$ near $q=i\infty$ \cite{cer12}: the conformal symmetry condition (\ref{toy})
fully determines the accessory parameters in the case $N=3$, the integration constants are determined by natural assumptions on the behaviour of $W_{m+1}(I,y)$ near $y=z_i$, and the Ward identities can be solved order by order in $q^2$ near $q=i\infty$ using Topological Recursion -- at each order, we have finitely many unknowns, and as many equations. We will now consider another special value of $q$, namely $q=0$, which corresponds to the central charge $c=1$. In this case, the differential terms $q\pp{y} W_{m+1}(I,y)$ of the Ward identities  vanish, and we will see that solutions can be constructed explicitly.

\subsection{Solving the \texorpdfstring{$c=1$}{c=1} Ward identities with matrices \label{ssecmat}}

Let us discuss the solutions of the Ward identities in the case $q=0$. By a solution we mean functions $Z_N,W_1,W_2,\cdots$ such that $W_m$ solve the Ward identities (\ref{qpw}), while $Z_N$ solves the global Ward identities (\ref{toy}). 
To find solutions, we will use a determinantal ansatz \cite{be09}. If $N\geq 3$, this ansatz leads to the construction of one or more solutions. In the case $N=3$ of Section \ref{sectp}, there actually exists only one solution up to a rescaling of $Z_N$ by $z_i$-independent factors, and this solution is therefore given by the ansatz. If $N\geq 4$, there are many solutions. As we will discuss in Section \ref{secps}, the ansatz does not give all of them -- but sufficiently many for the ansatz to be valuable. 

The ansatz depends on two matrices of size two: a Lax matrix $L(y)$, and a constant matrix $P$. 
The Lax matrix $L(y)$ is supposed to be traceless,
and to be a function of a spectral parameter $y$. The constant matrix $P$ is supposed to have the eigenvalues $1$ and $0$, so that $P^2=P$. 
We can now define two further matrices of size two: a matrix $\Psi(y)$ such that $\det \Psi(y)=1$ and 
\begin{align}
 \pp{y} \Psi(y) = L(y) \Psi(y) \ ,
\label{pplp}
\end{align}
and the following matrix $M(y)$,
\begin{align}
 M(y) = \Psi(y) P \Psi(y)^{-1} \quad \Rightarrow \quad \pp{y} M(y) = [L(y),M(y)]\ .
 \label{mp}
\end{align}
Using these matrices, we can write the determinantal ansatz,
\begin{align}
 W_1(y) &= -\mathrm{Tr} L(y)M(y)\ ,
\label{wo}
\\
W_2(y_1,y_2) & = -\frac{\mathrm{Tr}(M(y_1)-M(y_2))^2}{2(y_1-y_2)^2}\ , 
\label{wt}
\\
W_{m\geq 3}(y_1,\cdots y_m) & = \frac{(-1)^{m+1}}{m}\sum_{\sigma\in S_m}\frac{\mathrm{Tr}\prod_{i=1}^m M(y_{\sigma(i)})}{\prod_{i=1}^m(y_{\sigma(i)}-y_{\sigma(i+1)})}\ .
\label{wm}
\end{align} 
Inserting these $W_m$s in the left-hand side of the Ward identities (\ref{qpw}) yields values $P_{m+1}(I,y)$ of the right-hand side which were computed in \cite{be09}, and which we now require to agree with our definition (\ref{pm}) of $P_{m+1}(I,y)$. This requirement will result in constraints on the Lax matrix $L(y)$.

\paragraph{\fbox{Case $m=0$.}} 

Inserting the values (\ref{wo}) and (\ref{wt}) for $W_1$ and $W_2$ in eq. (\ref{qpw}) leads to 
\begin{align}
 P_1(y) = -W_2(y,y) -W_1(y)^2 = \frac12\mathrm{Tr} [L(y),M(y)]^2 - (\mathrm{Tr} L(y)M(y))^2\ .
\end{align}
Let us now use the characteristic equation of size two matrices,
\begin{align}
 A^2 -(\mathrm{Tr}A)A+(\det A)\mathrm{Id} = 0\ .
\label{aaa}
\end{align}
Applying this to $A=L(y)M(y)$ and remembering $\det M(y)=0$, we obtain $P_1(y)= -\mathrm{Tr} L(y)^2 M(y)^2$. Applying the characteristic equation to $A=L(y)$ and remembering $\mathrm{Tr} L(y)=0$ and $\mathrm{Tr} M(y)^2=1$, we then obtain  
\begin{align}
 P_1(y) = \det L(y)\ .
\end{align}
According to our expression (\ref{pm}) for $P_1(y)$, we must therefore have 
$\det L(y) = -t(y)$ where $t(y)$ is given by eq. (\ref{ty}). Using eq. (\ref{aaa}) and remembering $\mathrm{Tr}L(y)=0$, this implies that $L(y)$ must provide a matrix square root of the expectation value of the energy-momentum tensor $T(y)$,
\begin{align}
 \boxed{L(y)^2 = \lla T(y) \rra  \mathrm{Id}}\ .
\label{llt}
\end{align}
Assuming that $L(y)$ is a rational function of $y$, this leads to
\begin{align}
 L(y) = \sum_{i=1}^N \frac{L_i}{y-z_i}\ , \quad \mathrm{where} \quad \bla \mathrm{Tr} L_i = 0\ ,
 \\ \det L_i = -\Delta_i\ , \\ \sum_{i=1}^N L_i = 0\ , \ela
 \label{ly}
\end{align}
where the condition $\sum_{i=1}^N L_i=0$ comes from the conformal symmetry condition (\ref{toy}). The eigenvalues of $L_i$ are therefore $p_i$ and $-p_i$.

\paragraph{\fbox{Case $m=1$.}}

We do not reproduce the calculation of the value of $P_2(y_1,y)$ implied by the ansatz (\ref{wo})-(\ref{wm}) for $W_m$, and simply quote the result \cite{be09}
\begin{align}
 P_2(y_1,y) = \mathrm{Tr}\left(\frac{L(y)-L(y_1)-(y-y_1)L'(y_1)}{(y-y_1)^2} M(y_1)\right)\ .
 \label{ptt}
\end{align}
Now our expression (\ref{pm}) for $P_{m\geq 2}(I,y)$ involves derivatives $\pp{z_i}$. We should therefore know how the various matrices depend on $z_i$. We assume the constant matrix $P$ in eq. (\ref{mp}) to be independent not only from $y$, but also from $z_i$. We introduce matrices $R_i(y)$ such that 
\begin{align} 
 \pp{z_i} \Psi(y) = R_i(y) \Psi(y)\ .
 \label{pzp}
\end{align}
The matrices $R_i(y)$ are subject to the following compatibility conditions, which follow from $[\pp{y},\pp{z_i}]\Psi(y) = [\pp{z_i},\pp{z_j}]\Psi(y) = 0$, 
\begin{align}
 \pp{z_i} L(y) - \pp{y} R_i(y)  - [R_i(y),L(y)] &= 0 \ ,
 \label{plpr}
 \\
 \pp{z_j} R_i(y) - \pp{z_i} R_j(y) - [R_j(y),R_i(y)] &= 0\ .
 \label{pzpz}
\end{align}
Now eq. (\ref{ptt}) amounts to equations for the matrices $R_i(y)$, namely $\pp{y} R_i(y) = \frac{L_i}{(y-z_i)^2}$. The solution is
\begin{align}
 R_i(y) = -\frac{L_i}{y-z_i} + A_i \ ,
\label{rla}
\end{align}
where $A_i$ is some $y$-independent matrix. The compatibility conditions (\ref{plpr}) and (\ref{pzpz}) now boil down to
\begin{align}
 \pp{z_j} L_i -\frac{[L_i,L_j]}{z_i-z_j} - [A_j,L_i] & = 0\ , \quad (j\neq i)\ ,
\label{lij}
 \\
 \pp{z_i} L_i -\sum_{j\neq i}\frac{[L_j,L_i]}{z_i-z_j} - [A_i,L_i] & = 0 \ ,
\label{lii}
 \\
 \pp{z_j} A_i - \pp{z_i} A_j - [A_j,A_i] & = 0\ .
\label{aij}
\end{align}
The matrices $A_i$ are auxiliary objects, which can be set to zero by a change of unknown function $\Psi(y) \rightarrow \Theta\Psi(y)$. Then the equations (\ref{lij})-(\ref{lii}) reduce to Schlesinger's isomonodromy equations, which mean that the monodromies of $\Psi(y)$ are invariant under small changes of the positions $z_i$. This isomonodromy property was derived as a consequence of the Ward identities and therefore, ultimately, of local conformal symmetry. So isomonodromy for the variables $z_i$ expresses conformal symmetry at the level of the Lax matrix. (However, the very existence of the Lax matrix, in other words the validity of our ansatz, does not follow from conformal symmetry.)

Let us now derive the consequences of the equations (\ref{lij})-(\ref{aij}) for the Lax matrix $L(y)$. We view these equations as 
$L(y)$-dependent equations for $A_i$, and study the conditions for the existence of solutions. An equation of the type $[L_i,A]=E_i$ has a solution $A$ provided $\mathrm{Tr}L_iE_i=0$, and two equations $[L_i,A]=E_i$ and $[L_j,A]=E_j$ have a common solution $A$ provided $\mathrm{Tr}(L_iE_j+L_jE_i)=0$. So the equations (\ref{lij}) and (\ref{lii}) lead to the following constraints on the Lax matrix,
\begin{align}
 \pp{z_k}\mathrm{Tr} L_iL_j +\left(\frac{1}{z_i-z_k}-\frac{1}{z_j-z_k}\right)\mathrm{Tr}[L_i,L_j]L_k & = 0 \ , \quad (i\neq j\neq k)
\label{dkij}
\\
\pp{z_i} \mathrm{Tr} L_iL_j +\mathrm{Tr}[L_i,L_j]\bigg(\sum_{k\neq i,j}\frac{L_k}{z_k-z_i}\bigg) &= 0\ .
\label{dij}
\end{align}

\paragraph{\fbox{Case $m\geq 2$.}}

The ansatz (\ref{wo})-(\ref{wm}) for $W_m$ leads to \cite{be09}
\begin{align}
 P_{m+1}(I,y) &= Q_{m+1}(I,y)-\sum_{i=1}^m\frac{1}{y-y_i}\underset{y'=y_i}{\mathrm{Res}} Q_{m+1}(I,y')\ ,
\\
 \mathrm{with} \quad Q_{m+1}(I,y) &= (-1)^m\sum_{\sigma\in S_m} \frac{\mathrm{Tr} L(y) \prod_{i=1}^m M(x_{\sigma(i)})}{(y-y_{\sigma(1)}) \prod_{i=1}^{m-1}(y_{\sigma(i)}-y_{\sigma(i+1)})\ (y_{\sigma(m)}-y)}\ . 
\end{align}
The analytic properties of this $P_{m+1}(I,y)$ as a function of $y$ agree with eq. (\ref{pm}): we have poles at $y=z_i$, and vanishing at $y=\infty$. It remains to be checked that the residues at $y=z_i$ agree. Using eqs. (\ref{wm}), (\ref{pzp}) and (\ref{rla}) we find 
\begin{align}
 \pp{z_i} W_m(I) =  \sum_{\sigma\in S_m} \frac{(-1)^{m}\mathrm{Tr} L_i \prod_{j=1}^m M(x_{\sigma(j)})}{(z_i-y_{\sigma(1)}) \prod_{j=1}^{m-1}(y_{\sigma(j)}-y_{\sigma(j+1)})\ (y_{\sigma(m)}-z_i)} = \underset{y=z_i}{\mathrm{Res}} P_{m+1}(I,y)\ .
\end{align}
Therefore, the case $m\geq 2$ does not lead to further constraints on the Lax matrix $L(y)$, besides the constraints (\ref{ly}) from the case $m=0$ and (\ref{dkij})-(\ref{dij}) from the case $m=1$.

\section{Four-point conformal blocks and the Painlev\'e VI equation \label{secps}}

This section is mostly a review of known results on how the Painlev\'e VI equation follows from the isomonodromy condition of Section \ref{ssecmat} in the case $N=4$, and how solutions of that equation are consequently related to conformal blocks. We include this material in order to complete the derivation of the Painlev\'e VI equation from the basic principles of Conformal Field Theory, and to discuss the interpretation of the solutions of that equation in terms of conformal blocks.

\subsection{Derivation of the Painlev\'e VI equation}

The $N$-point function $Z_N$, and the associated correlation functions $W_m$, are subject to the Ward identities, which are however expected to have many more solutions if $N\geq 4$. Actually, the Ward identities involve holomorphic variables $z_i$ and not their complex conjugates $\bar{z}_i$, and would also hold if we replaced $Z_N$ with a conformal block 
$\mathcal{F}_{p_s}(p_i|z_i)$ (\ref{zfi}) -- or more generally, with any linear combination of conformal blocks with $z_i$-independent coefficients. On the other hand, we have used a particular ansatz in Section \ref{ssecmat}, which is not guaranteed to yield all linear combinations of conformal blocks, or $Z_N$ itself. In an isomonodromic matrix differential equation such as eq. (\ref{pplp}), the object which plays the role of $Z_N$ is called the tau function. So we call 
$\tau_N$ a solution of the Ward identities which is given by our ansatz. With such a solution,
the accessory parameters 
\begin{align}
 \beta_i = \frac{1}{2\pi i}\oint_{z_i} t(y)dy = \frac{1}{2\pi i}\oint_{z_i}\frac12 \mathrm{Tr} L(y)^2 =\sum_{j\neq i}\frac{\mathrm{Tr}L_iL_j}{z_i-z_j}\ ,
\label{bll}
\end{align}
are no longer given by $\beta_i = \pp{z_i} \log Z_N$ as in eq. (\ref{ty}), but rather by
\begin{align}
 \beta_i = \pp{z_i}\log \tau_N\ .
\end{align}
Knowing the Lax matrix, and therefore $\beta_i$, determines $\tau_N$ up to $z_i$-independent factors, and we will neglect such factors in this Section.

$N$-point functions with $N\leq 3$ are completely determined by the conformal symmetry condition (\ref{toy}). We will therefore focus on the first non-trivial case $N=4$. Let us analyze the consequences for $\tau_4$ of the constraints (\ref{ly}) on the Lax matrix. 
The conformal symmetry condition (\ref{toy}) determines all but one accessory parameter, which can be chosen as
\begin{align}
 \sigma = \frac12\left[ z\mathrm{Tr}(L_2+L_3)^2 +(z-1)\mathrm{Tr}(L_1+L_2)^2\right]\ , \quad \mathrm{with} \quad z = \frac{z_{12} z_{34}}{z_{13}z_{24}}\ .
\label{sz}
\end{align}
 The relation between the accessory parameter $\sigma$ and $\tau_4$ is 
\begin{align}
 \sigma = z(z-1)\pp{z} \log\left(z^{\Delta_1+\Delta_2}(z-1)^{\Delta_2+\Delta_3} \tilde{\tau}_4\right)\ ,
\label{sig}
\end{align}
where $\tilde{\tau}_4$, which depends on $z_i$ only through the cross-ratio $z$, is defined by 
\begin{align}
 \tilde{\tau}_4 =
\left[\frac{z_{13}z_{24}^2}{z_{14}z_{34}}\right]^{\Delta_2} z_{13}^{\Delta_1+\Delta_3-\Delta_4} z_{34}^{\Delta_3+\Delta_4-\Delta_1} z_{14}^{\Delta_1+\Delta_4-\Delta_3}\tau_4 = \underset{(z_1,z_2,z_3,z_4)\rightarrow (0,z,1,\infty)}{\lim} z_4^{2\Delta_4} \tau_4\ .
\label{zzt}
\end{align}
Notice that $\mathrm{Tr}L_iL_j$ is also a function of the cross-ratio $z$, equivalently
\begin{align}
 D_y \mathrm{Tr} L_iL_j = O\left(\frac{1}{y^4}\right)\ ,
\label{dyll}
\end{align}
where $D_y$ is defined in eq. (\ref{dy}). This equation is proved by applying $\pp{z_i}$ to the conformal symmetry condition (\ref{toy}), and using $\pp{z_i} \beta_j=\pp{z_j}\beta_i$, which yields $(D_y+\frac{1}{(y-z_i)^2})\beta_i +\frac{2\Delta_i}{(y-z_i)^3} = O(\frac{1}{y^4})$. Then, using the expression (\ref{bll}) for $\beta_i$ as well as $\sum_i L_i=0$, we find $\sum_{j\neq i} \frac{D_y\mathrm{Tr}L_iL_j}{z_i-z_j} = O(\frac{1}{y^4})$. As this is true for any choice of $i$, and only two of the objects $\mathrm{Tr} L_iL_j$ are independent if $N=4$, we indeed obtain eq. (\ref{dyll}). (This equation can surely be assumed to hold for any value of $N$, although our argument does not fully prove it.) 

In the case $N=4$, the constraints (\ref{dkij})-(\ref{dij}) on the Lax matrix boil down to 
\begin{align}
 z\pp{z}\mathrm{Tr}L_2L_3 = z(z-1)\pp{z}\mathrm{Tr}L_1L_3 = (1-z)\pp{z}\mathrm{Tr}L_1L_2=\mathrm{Tr}L_1[L_2,L_3]\ ,
\label{tddd}
\end{align}
which in particular allows us to compute $\sigma'=\pp{z}\sigma$,
\begin{align}
 \sigma' = \frac12\left[ \mathrm{Tr}(L_2+L_3)^2 +\mathrm{Tr}(L_1+L_2)^2\right]\ .
\end{align}
The accessory parameter $\sigma$ and its derivative $\sigma'$ can actually be considered as the two ``matrix accessory parameters'' -- the parameters of the Lax matrix $L(y)$ which are left undetermined by the constraints (\ref{ly}), and which are invariant under the transformations
\begin{align}
L(y)\mapsto \Lambda L(y) \Lambda^{-1}\ .
\label{lll}
\end{align}
All terms in eq. (\ref{tddd}) can therefore be rewritten in terms of $\sigma$ and $\sigma'$. In the case of the last, cubic term, this is done with the help of the following identity for traceless matrices of size two,
\begin{align}
 \left(\mathrm{Tr}ABC\right)^2 = -\frac12 \det\begin{pmatrix} \mathrm{Tr} A^2 & \mathrm{Tr} AB & \mathrm{Tr} AC \\ \mathrm{Tr} AB & \mathrm{Tr} B^2 & \mathrm{Tr} BC \\ \mathrm{Tr} AC & \mathrm{Tr} BC & \mathrm{Tr} C^2 \end{pmatrix}\ .
\end{align}
We thus obtain a differential equation for the accessory parameter $\sigma$,
\begin{align}
 \left(z(z-1)\sigma''\right)^2 & = -2\det \begin{pmatrix} 2\Delta_1 & Z  & Y \\ Z & 2\Delta_2 & X \\ Y & X & 2\Delta_3 \end{pmatrix}, \
\mathrm{where}\ \bla 
X=(1-z)\sigma'+\sigma-\Delta_2-\Delta_3\ , \\ Y = -\sigma'+\Delta_2+\Delta_4\ , \\ Z = z\sigma'-\sigma-\Delta_1-\Delta_2\ . \ela 
\label{sps}
\end{align}
This equation is called the sigma form of the Painlev\'e VI equation. Via eq. (\ref{sig}), it amounts to a differential equation for $\tau_4$, which can now be identified as the tau function of the Painlev\'e VI equation.
This equation, its relation to the matrix differential equation (\ref{pplp}), and its relevance to Conformal Field Theory with $c=1$, are not new \cite{gil12}. What we hope to have achieved is a derivation of the sigma-Painlev\'e VI equation from minimal assumptions in Conformal Field Theory. In particular, discussing the field content (in other words the spectrum) of the theory, as was done in \cite{gil12}, is not necessary.

\subsection{Solutions of the Painlev\'e VI equation}

Let us discuss the solutions of the sigma-Painlev\'e VI equation and their interpretation, reviewing and commenting some results of \cite{gil12,ilt13}. 

Let us first discuss whether the tau function $\tau_4$ can coincide with an $s$-channel conformal block $\mathcal{F}_{p_s}(p_i|z_i)$. 
The function $\tilde{\mathcal{F}}_{p_s}(p_i|z)=\underset{(z_1,z_2,z_3,z_4)\rightarrow (0,z,1,\infty)}{\lim} z_4^{2\Delta_4}\mathcal{F}_{p_s}(p_i|z_i)$ of the cross-ratio $z$, which corresponds to the conformal block $\mathcal{F}_{p_s}(p_i|z_i)$ through a relation similar to eq. (\ref{zzt}), behaves near $z=0$ as 
\begin{align}
 \tilde{\mathcal{F}}_{p_s}(p_i|z) = z^{\Delta_s-\Delta_1-\Delta_2}\left(1+ \sum_{k=1}^\infty c_k z^k\right)\ ,
\end{align}
where
$c_k$ is a $z$-independent coefficient which is a function of $\Delta_s$ and $\Delta_i$. As a consequence, the corresponding function $\sigma^\mathrm{block}$ (\ref{sig})
would be analytic near $z=0$, with the behaviour
$
 \sigma^\mathrm{block} = -\Delta_s + \sum_{k=1}^\infty \sigma_k z^k .
$
However, expanding the sigma-Painlev\'e VI equation (\ref{sps}) near $z=0$ shows that, for given values of $\Delta_i$, such an analytic solution can exist only for four particular values of $\Delta_s$. Writing $\Delta=p^2$ as in eq. (\ref{da}), the corresponding eight values of the momentum $p_s$ are 
\begin{align}
 p_s = \pm p_1\pm p_2\quad , \quad p_s = \pm p_3\pm p_4\ .
\end{align}
This is enough for describing the rather trivial conformal blocks of free bosonic theories. In such theories the momentum $p$ is conserved, and under the assumption $\sum_{i=1}^4 p_i =0$ we find the simple solution
\begin{align}
 \sigma^\mathrm{free} = (p_2+p_3)^2z + (p_1+p_2)^2(z-1)\ .
\end{align}

Generic conformal blocks exist for arbitrary values of $\Delta_s$, and are not solutions of the Painlev\'e VI equation. 
Rather, solutions can be built from infinite families of conformal blocks, whose momentums belong to $p_s+\Z$ for a given $p_s$. This is natural from the point of view of the matrix differential equation (\ref{pplp}), whose monodromy matrix around a contour enclosing both $z_1$ and $z_2$ has eigenvalues $e^{\pm 2\pi ip_s}$ -- loosely speaking, the matrix differential equation knows $p_s$ only modulo integers. So, for any choice of $p_s$, and of a value for a new parameter $x$, we have the solution  \cite{ilt13}
\begin{align}
 \tau_4 = \sum_{n\in \Z} C_{p_s+n}(p_i) x^n \mathcal{F}_{p_s+n}(p_i|z_i)\ ,
\label{tf}
\end{align}
where the coefficient $C_{p_s}(p_i)$ is given in terms of Barnes' $G$-function by
\begin{align}
 C_{p_s}(p_i) = \frac{\prod_{\epsilon,\epsilon'=\pm} G(1 +\epsilon p_s+p_2+\epsilon' p_1) G(1+\epsilon p_s+p_3+\epsilon'p_4)}{\prod_{\epsilon=\pm} G(1+2\epsilon p_s)} \ ,
\end{align}
and $\tau_4$ only depends on $p_s$ modulo integers in the sense that $\left.\tau_4\right|_{p_s\rightarrow p_s+n} = x^{-n}\tau_4$.
The family of functions $(\tau_4)_{\{p_s\}\in (0,1),x\in \mathbb{C}}$ forms a basis of solutions of the Ward identities. This basis is related to the basis $(\mathcal{F}_{p_s}(p_i|z_i))_{p_s\in \mathbb{R}}$ by eq. (\ref{tf}), which can be thought of as the expansion of $\tau_4$ as a Laurent series in $x$. This Laurent expansion maps the dependence of $\mathcal{F}_{p_s}(p_i|z_i)$ on the integer part of $p_s$ to the dependence of $\tau_4$ on $x$. 
The inverse change of basis is given by $\mathcal{F}_{p_s}(p_i|z_i) = \frac{1}{2\pi i C_{p_s}(p_i)}\oint_0 \frac{dx}{x} \tau_4$.

The decomposition (\ref{tf}) of the tau function $\tau_4$ of the Painlev\'{e} VI equation into conformal blocks can now be compared to the decomposition (\ref{zfi}) of the four-point function $Z_4$. The formal similarity of these decompositions may suggest that $\tau_4$ can be interpreted as the four-point function of some theory. (The theory would live on the disc rather than on the complex plane, since we have conformal blocks rather than their moduli squared.) This interpretation is however incorrect, for a number of reasons:
\begin{enumerate}
 \item The tentative spectrum of the theory would be characterized by the set of conformal dimensions $\Delta\in (p_s+\Z)^2$. But the corresponding set of representations of the Virasoro algebra is not closed under fusion.
\item $\tau_4$ is not a single-valued function of $z_i$. 
\item In the case of a free bosonic theory, $\tau_4$ is a conformal block, not a correlation function.
\item The coefficient $C_{p_s}(p_i)x^n$ does not factorize into a product of structure constants \cite{gil12}, as would happen in the decomposition of a correlation function. Notice however that the coefficient $C_{p_s}(p_i)$ is built from the same special function $G$ as the three-point function (\ref{cco}) of Runkel-Watts theory.
\end{enumerate}

To summarize, the solutions of the Ward identities which are built from our ansatz are the tau functions $\tau_4$ of the Painlev\'e VI equation. While they include neither the conformal blocks $\mathcal{F}_{p_s}(p_i|z_i)$ nor the four-point function $Z_4$, these tau functions form a basis of the space of solutions. This provides a way to compute conformal blocks by solving the sigma-Painlev\'e VI equation, or conversely to compute the Painlev\'e VI tau function $\tau_4$ from conformal blocks. 

\section{The three-point function of Runkel-Watts theory \label{sectp}}

In Section \ref{secps} we studied correlation functions $Z_N$ up to $z_i$-independent factors. But such factors play a crucial role in Conformal Field Theory. In particular, determining the three-point structure constant in principle amounts to solving the model under consideration. So in this Section we will focus on the case $N=3$, even though the dependence of $Z_3$ on $z_i$ is trivial. 
To access $z_i$-independent factors, we will need Seiberg-Witten equations. Such equations were introduced in \cite{cer12} as axioms; here we will derive them from the principles of Conformal Field Theory.

\subsection{Seiberg-Witten equations \label{ssecsw}}

Coming back to the formalism of Section \ref{ssecwi}, let us make the standard assumption that the energy-momentum tensor $T(z)$ generates the conformal transformations, and in particular
\begin{align}
\pp{z} V_p(z,\bar{z})= \frac{1}{2\pi i} \oint_z dy\ T(y)V_p(z,\bar{z}) \ .
\label{itv}
\end{align}
Using Wick's theorem (see for instance \cite{bs92}) for deducing the singular terms of $T(y)V_p(z,\bar{z})$ from eqs. (\ref{jv}) and (\ref{tj}), we obtain
\begin{align}
 \pp{z} V_p(z,\bar{z}) = 2p (JV_p)(z,\bar{z})\ .
\label{pvjv}
\end{align}
Neglecting issues of regularization, we may rewrite this equation as $\nabla_z V_p(z,\bar{z}) =0 $ where $\nabla_z = \pp{z}-2p J(z)$. This suggests that we introduce $\nabla_p = \pp{p} -2\int^z dy J(y)$, such that $[\nabla_z,\nabla_p]=0$. We therefore have $\nabla_z \nabla_p V_p(z,\bar{z})=0$, thus $\nabla_p V_p(z,\bar{z})$ belongs to the kernel of $\nabla_z$. This implies $\nabla_p  V_p(z,\bar{z}) = \lambda(p) V_p(z,\bar{z})$ for some function $\lambda(p)$, which we can however set to zero by renormalizing the field $V_p(z,\bar{z})$. We obtain the Seiberg-Witten equation
\begin{align}
 \pp{p}V_p(z,\bar{z}) = 2\int^z dy\ J(y)V_p(z,\bar{z})\ .
\label{sw}
\end{align}
In terms of correlation functions, this amounts to 
\begin{align}
 \pp{p_i}\log Z_N = 2 \int^{z_i} dy\ W_1(y) \ ,
\label{paz}
\end{align}
which holds under the assumption, which follows from eq. (\ref{jv}), 
\begin{align}
 W_1(y) \underset{y\rightarrow z_i}{=} \frac{p_i}{y-z_i} + O(1)\ .
\label{wizi}
\end{align}
To make precise sense of eq. (\ref{paz}), we have to specify a second integration bound, and to deal with the divergence of the integral at $y=z_i$. It would be tempting to regularize the integral as $\int^{z_i}dy\left(W_1(y)-\frac{p_i}{y-z_i}\right)$. But then the contribution of the subtracted term would depend on the choice of the second integration bound. Rather, we should use a "local" regularization, which does not influence the integral away from the singularity. This is provided by 
\begin{align}
 \int^{z_i} dy\ W_1(y)\ \underset{\mathrm{regularized}}{=}\ \underset{\epsilon_i\rightarrow 0}{\lim} \left(\frac{p_i}{\epsilon_i} + \int^{z_i} dy\ (y-z_i)^{\epsilon_i} W_1(y) \right)\ .
\label{ireg}
\end{align}
For the second integration bound, we propose to take another singularity $z_j$, so that the Seiberg-Witten equation becomes 
\begin{align}
 \left(\pp{p_i}-\pp{p_j}\right)\log Z_N = 2\int^{z_i}_{z_j} dy\ W_1(y)\ .
\label{ppaz}
\end{align}
Let us recapitulate the conditions under which this equation is supposed to hold:
\begin{enumerate}
 \item the behaviour of $W_1(y)$ near both singularities $z_i$ and $z_j$ is determined by eq. (\ref{wizi}), 
\item the regularization (\ref{ireg}) is implied.
\end{enumerate}
Furthermore, we have the caveat that this 
equation, which was obtained by integrating a $z_i$-differential equation, does not constrain the dependence of $Z_N$ on $\bar{z}_i$.

\subsection{The Lax matrix and the associated equation}

In the case $N=3$, the Lax matrix $L(y)$ is determined by the conditions (\ref{ly}) up to the conjugation (\ref{lll}) by a constant matrix, which can a priori depend on $z_i$. Actually, we can make the assumption
that the matrices $L_j$ are independent not only from $y$, but also from $z_i$. This is because, under this assumption, the compatibility equations (\ref{lij})-(\ref{aij}) have the solution
\begin{align}
 A_i =\frac{L_j}{z_i-z_k} + \frac{L_k}{z_i-z_j}\ , \quad \mathrm{where} \quad i\neq j\neq k\ .
\end{align}
This determines the matrices $R_i(y)$ in terms of the constant matrices $L_j$. Let us now study $\Psi(y)$, as defined by 
the equations (\ref{pplp}) and (\ref{pzp}). From these equations, the behaviour of $\Psi(y)$ near the singular point $z_i$ is 
\begin{align}
 \Psi(y) \underset{y\rightarrow z_i}{=} \left(\mathrm{Id} + O\left(y-z_i\right)\right) Y_i^{L_i} C_i\quad \mathrm{with} \quad Y_i =
\frac{(y-z_i)(z_k-z_j)}{(z_i-z_k)(z_i-z_j)}\ ,
\label{plc}
\end{align}
where $C_i$ is a constant matrix of determinant one. Choosing $C_i$ for some index $i$ determines a solution $\Psi(y)$.

From now on we omit the $y$-dependence in objects such as $\Psi=\Psi(y)$, while denoting their $y$-derivatives with a prime.
Let us introduce notations for the matrix elements of $\Psi$:
\begin{align}
 \Psi = K \hat{\Psi}\ ,\quad \mathrm {where} \quad  \hat{\Psi} =\begin{pmatrix} \psi_+ & \psi_- \\ \psi_+' & \psi_-' \end{pmatrix}  \ \mathrm{and}\ \ K = \begin{pmatrix} 1 &  0 \\ -\frac{a}{b} & \frac{1}{b} \end{pmatrix} ,\ \mathrm{if}\ \ L = \begin{pmatrix} a & b \\ c & d \end{pmatrix}\ .
\label{pkp}
\end{align}
The functions $\psi_\pm$ obey the differential equation 
\begin{align}
 \psi_i'' -\frac{b'}{b}\psi_i' + \left(\frac{ab'-a'b}{b} - t\right) \psi_i = 0 \ .
\label{ppi}
\end{align}
Then $W_1$ (\ref{wo}) can be rewritten as 
\begin{align}
 W_1 = -\mathrm{Tr} P \Psi^{-1} L \Psi\ , 
\label{wplp}
\end{align}
where 
\begin{align}
\Psi^{-1} L \Psi = \hat{\Psi}^{-1} \begin{pmatrix} 0 & 1 \\ t & 0 \end{pmatrix} \hat{\Psi} 
= \frac{1}{b}\begin{pmatrix} \psi_+'\psi_-'-t\psi_+\psi_- & (\psi_-')^2 - t(\psi_-)^2 \\  -(\psi_+')^2+t(\psi_+)^2 & -\psi_+'\psi_-'+t\psi_+\psi_- \end{pmatrix}
\ .
\label{wpt}
\end{align}
The matrix elements of $\Psi^{-1}L\Psi$ are of the type 
\begin{align}
 \frac{\psi_i'\psi_j'-t\psi_i\psi_j}{b} = \left(\frac{\psi_i\psi_j'}{b}\right)' +\frac{\psi_i\psi_j}{b}\left(\frac{ab'-a'b}{b}-2t\right)\ .
\label{ptp}
\end{align}
(This relation holds for any pair $(\psi_i,\psi_j)$ of solutions of eq. (\ref{ppi}).)

Let us explicitly compute $L$ and $\Psi$. Assuming that $L_1$ is diagonal, we find 
\begin{align}
 L_1 & = \begin{pmatrix} p_1 & 0 \\ 0 & -p_1 \end{pmatrix} \ ,
\label{lo}
\\
L_2 & = \begin{pmatrix} -\frac{\Delta_1+\Delta_2-\Delta_3}{2p_1} & -B \\ -C & \frac{\Delta_1+\Delta_2-\Delta_3}{2p_1} \end{pmatrix}\ ,
\label{loo}
\\
L_3 & = \begin{pmatrix} -\frac{\Delta_1-\Delta_2+\Delta_3}{2p_1} & B \\ C & \frac{\Delta_1-\Delta_2+\Delta_3}{2p_1} \end{pmatrix}  \ ,
\label{looo}
\end{align}
where $B$ and $C$ are numbers such that 
\begin{align}
 BC = \frac{2\Delta_1\Delta_2+2\Delta_2\Delta_3+2\Delta_3\Delta_1-\Delta_1^2-\Delta_2^2-\Delta_3^2}{4\Delta_1}\ .
\end{align}
The differential equation (\ref{ppi}) for the coefficients of the matrix $\Psi$ now becomes the hypergeometric differential equation. 
Let us call $\Psi^{(1)}$ the solution whose 
matrix $C_1$ in eq. (\ref{plc}) is $C_1 = \mathrm{Id}$. The matrix elements of $\Psi^{(1)}$, as defined in eq. (\ref{pkp}), are 
\begin{align}
 \psi^{(1)}_+ &= Y^{p_1} (1-Y)^{p_2}F(p_1+p_2+p_3,p_1+p_2-p_3,2p_1,Y)\ , 
\label{pop}
\\
\psi^{(1)}_- &= \frac{B}{(1-2p_1)} Y^{1-p_1}(1-Y)^{p_2}F(p_2+p_3-p_1+1,p_2-p_3-p_1+1,2-2p_1,Y)\ ,
\label{pom}
\end{align}
where $F(a,b,c,Y)$ is the hypergeometric function, and
we introduced 
\begin{align}
 Y = \frac{(y-z_1)(z_3-z_2)}{(y-z_3)(z_1-z_2)}\ .
\label{yy}
\end{align}
We also define a solution $\Psi^{(2)}$ by 
\begin{align}
 \psi^{(2)}_+ &= Y^{p_1}(1-Y)^{p_2}F(p_1+p_2+p_3,p_1+p_2-p_3,2p_2+1,1-Y)\ ,
\label{poop}
\\
 \psi^{(2)}_-&= \frac{B}{2p_2} Y^{p_1}(1-Y)^{-p_2}F(p_1-p_2+p_3,p_1-p_2-p_3,1-2p_2,1-Y)\ .
\label{poom}
\end{align}
There must exist a constant matrix $F$ of determinant one such that 
\begin{align}
 \Psi^{(1)} =  \Psi^{(2)}F\ .
\label{ppf}
\end{align}
Explicitly, we find 
\begin{align}
 F = \begin{pmatrix} f_{++} & f_{+-} \\ f_{-+} & f_{--} \end{pmatrix} 
= \begin{pmatrix} \frac{\Gamma(2p_1)\Gamma(-2p_2)}{\Gamma(p_1-p_2-p_3)\Gamma(p_1-p_2+p_3)} & B \frac{\Gamma(1-2p_1)\Gamma(-2p_2)}{\Gamma(1-p_1-p_2-p_3)\Gamma(1-p_1-p_2+p_3)} \\  B^{-1}\frac{\Gamma(2p_1)\Gamma(1+2p_2)}{\Gamma(p_1+p_2+p_3)\Gamma(p_1+p_2-p_3)} 
& \frac{\Gamma(1-2p_1)\Gamma(1+2p_2)}{\Gamma(1-p_1+p_2+p_3)\Gamma(1-p_1+p_2-p_3)} \end{pmatrix}\ .
\label{fmat}
\end{align}
We have chosen the solutions $\Psi^{(1)}$ and $\Psi^{(2)}$ so that the associated  $W_1$ functions (\ref{wplp}) behave simply near $y=z_1$ and $y=z_2$ respectively. 
From our definition of $\Psi^{(1)}$, we have 
\begin{align}
 (\Psi^{(1)})^{-1}L\Psi^{(1)} \underset{y\rightarrow z_1}{=} \frac{\left(\begin{smallmatrix} p_1 & 0 \\ 0 &  -p_1\end{smallmatrix}\right)}{y-z_1} + O(1)\ .
\label{plpo}
\end{align}
With the help of eq. (\ref{wpt}), we moreover find 
\begin{align}
 (\Psi^{(2)})^{-1}L\Psi^{(2)} \underset{y\rightarrow z_2}{=} \frac{\left(\begin{smallmatrix} p_2 & 0 \\ 0 & -p_2\end{smallmatrix}\right)}{y-z_2} + O(1)\ .
\label{plpt}
\end{align}

Having computed the Lax matrix $L$, let us complete the determination of the ansatz of Section \ref{ssecmat} for the $W_m$s by computing the constant matrix $P$. The choice of $P$ is constrained by our intention to use the Seiberg-Witten equation (\ref{ppaz}) for computing the corresponding partition function $Z_3$. For that equation to hold, 
we have to assume that $W_1$ obeys eq. (\ref{wizi}) near two singularities, which we choose to be $z_1$ and $z_2$. 
The behaviour of $W_1$ near $z_1$ and $z_2$ is determined by the equations (\ref{wplp}), (\ref{ppf}), (\ref{plpo}) and (\ref{plpt}). Working with the solution $\Psi^{(1)}$ so that $W_1 = -\mathrm{Tr} P \left(\Psi^{(1)}\right)^{-1}L\Psi^{(1)}$, we obtain the following constraints on the matrix $P$:
\begin{align}
 \mathrm{Tr} P \left(\begin{smallmatrix} 1 & 0 \\ 0 & -1 \end{smallmatrix}\right) = \mathrm{Tr} FPF^{-1} \left(\begin{smallmatrix} 1 & 0 \\ 0 & -1 \end{smallmatrix}\right) = -1 \ .
\end{align}
If we remember that $P$ is supposed to have the eigenvalues $1$ and $0$, this leads to two solutions for the matrix $P$, which can be expressed in terms of the elements of the matrix $F$ (\ref{fmat}),
\begin{align}
 P^+ = \begin{pmatrix} 0 & 0 \\ \frac{f_{-+}}{f_{--}} & 1 \end{pmatrix}  \quad \mathrm{and} \quad P^- =\begin{pmatrix} 0 & -\frac{f_{+-}}{f_{++}} \\ 0 & 1 \end{pmatrix}\ .
\end{align}
Which one of these two solutions should we use? Calling $W_1^\pm$ the $W_1$ function built from $P^\pm$, the correct prescription turns out to be 
\begin{align}
 W_1=\frac12(W_1^++W_1^-)\ ,
\label{wopm}
\end{align}
so that the Seiberg-Witten equation becomes
\begin{align}
 \left(\pp{p_1}-\pp{p_2}\right)\log Z_3 = \int_{z_2}^{z_1} dy\left(W_1^+(y)+W_1^-(y) \right)\ ,
\label{ppzt}
\end{align}
where the regularization (\ref{ireg}) is implied.
What singles out our prescription for $W_1$ is making the right-hand side of this equation antisymmetric under the permutation $1\leftrightarrow 2$, as we will see.

\subsection{Computation of the three-point function}

To perform the integral (\ref{ppzt}), we first write the integrand as 
\begin{align}
 W_1^+ + W_1^- = \frac{\left(\psi_+^{(1)}\right)'\left(\psi_-^{(2)}\right)'-t\psi_+^{(1)}\psi_-^{(2)}}{bf_{++}} + \frac{\left(\psi_-^{(1)}\right)'\left(\psi_+^{(2)}\right)'-t\psi_-^{(1)}\psi_+^{(2)}}{bf_{--}}\ ,
\label{wpwm}
\end{align}
which follows from eq. (\ref{wpt}) and the identity $F(P^++P^--\mathrm{Id}) = \left(\begin{smallmatrix} -\frac{1}{f_{--}} & 0 \\ 0 & \frac{1}{f_{++}} \end{smallmatrix}\right)$. Let us split the integral into three terms,
\begin{align}
 \left(\pp{p_1}-\pp{p_2}\right)\log Z_3= I_1+I_2+I_3\ ,
\end{align}
which we now define and compute. The first term $I_1$ is the contribution of the first, total derivative term of eq. (\ref{ptp}) when applied to eq. (\ref{wpwm}), so that
\begin{align}
 I_1 = \left[\frac{\psi_+^{(1)}\left(\psi_-^{(2)}\right)'}{bf_{++}}+\frac{\psi_-^{(1)}\left(\psi_+^{(2)}\right)'}{bf_{--}}\right]_{z_2}^{z_1} = \left(\frac{\left(\psi_-^{(1)}\psi_+^{(1)}\right)'}{b}\right)(z_1)-\left(\frac{\left(\psi_-^{(2)}\psi_+^{(2)}\right)'}{b}\right)(z_2)\ ,
\end{align}
where we use the relation (\ref{ppf}) between the solutions $\Psi^{(1)}$ and $\Psi^{(2)}$ so as to keep only the terms which have finite limits at $z_1$ and $z_2$. (The rest of the terms are eliminated by our regularization (\ref{ireg}) under suitable assumptions on the momentums $p_i$.) Explicitly, we find
\begin{align}
 I_1 = -\frac{1}{2p_1-1}\ .
\end{align}
Let us now compute the contribution of the second term in eq. (\ref{ptp}). We split this contribution into two terms $I_2+I_3$ in order to confine the dependence on $z_i$ into one simple term, namely $I_2$. It may seem at first sight that all $z_i$-dependence can be eliminated from eq. (\ref{ppzt}) by changing the integration variable from $y$ to $Y$ (\ref{yy}). This is however spoiled by the regularizing factors (\ref{ireg}). We would have a $z_i$-independent integral if we used the regularization based on the factor $Y^{\epsilon_1}(1-Y)^{\epsilon_2}$ instead of $(y-z_1)^{\epsilon_1}(y-z_2)^{\epsilon_2}$. We call $I_2$ the difference between these two regularizations. To compute $I_2$, we only need to know the behaviour (\ref{wizi}) of the integrand near the singularities $z_1$ and $z_2$. We find 
\begin{align}
 I_2 = -2p_1 \log \frac{(z_1-z_3)(z_1-z_2)}{z_2-z_3} +2p_2 \log \frac{(z_2-z_1)(z_2-z_3)}{z_1-z_3}\ .
\label{ioo}
\end{align}
The remaining contribution to $\left(\pp{p_1}-\pp{p_2}\right)\log Z_3$ is 
\begin{align}
 I_3 = 2 \int_0^1 \frac{dY}{Y} \left(\frac{\psi_+^{(1)}\psi_-^{(2)}}{Bf_{++}}+\frac{\psi_-^{(1)}\psi_+^{(2)}}{Bf_{--}}\right)\left(\frac{p_1(p_1-\frac12)}{Y} + \frac{p^2_2}{1-Y} - p^2_3\right)\ ,
\label{it}
\end{align}
where the implicit regularization is now 
\begin{align}
 \int_0^1 \frac{dY}{Y}\cdots \ \underset{\mathrm{regularized}}{=}\ \underset{\epsilon_{1,2}\rightarrow 0}{\lim}\left[ \frac{p_1}{\epsilon_1}-\frac{p_2}{\epsilon_2} + \int_0^1 \frac{dY}{Y} Y^{\epsilon_1}(1-Y)^{\epsilon_2} \cdots \right]\ .
\end{align}
The $\psi_+^{(1)}\psi_-^{(2)}$ term of $I_3$ can be written as $I(p_1-p_2+p_3,p_1-p_2-p_3,2p_1)$, if we define
\begin{multline}
 I(a,b,c) = \frac12 \frac{\Gamma(a)\Gamma(b)}{\Gamma(c)\Gamma(a+b-c+1)} \int_0^1 dY\ Y^{c-1}(1-Y)^{a+b-c}
\\ \left(-\frac{c(c-1)}{Y}+\frac{(a+b-c)^2}{Y-1}+(a-b)^2\right) F(a,b,c,Y)F(a,b,a+b-c+1,1-Y)\ . 
\end{multline}
Expanding both hypergeometric factors into series, we find 
\begin{align}
 I(a,b,c) = a\psi(a)+b\psi(b)-(a+b-c)\psi(a+b-c)-c\psi(c)+\frac{1}{2(c-1)}\ ,
\end{align}
where we use the digamma function
\begin{align}
 \psi(x) = \frac{\Gamma'(x)}{\Gamma(x)}\ .
\end{align}
(For details on the calculation of very similar integrals, see Appendix A.3 of \cite{cer12}.) The $\psi_-^{(1)}\psi_+^{(2)}$ term of $I_3$ is $-I(1-p_1+p_2+p_3,1-p_1+p_2-p_3,2-2p_1)$, up to a small discrepancy in the $-\frac{c(c-1)}{Y}$ term of $I(a,b,c)$. Taking this discrepancy into account, we obtain the following expression for the $z_i$-independent terms of $\left(\pp{p_1}-\pp{p_2}\right)\log Z_3$,
\begin{align}
 I_1+I_3 = \tilde{\psi}(2p_2)-\tilde{\psi}(2p_1) + \tilde{\psi}(p_1-p_2+p_3)-\tilde{\psi}(-p_1+p_2+p_3)\ ,
\label{ioit}
\end{align}
where we defined
\begin{align}
 \tilde{\psi}(x) = x\left(\psi(x)+\psi(-x)\right)\ .
\end{align}
Our expression for $I_1+I_3$ is now manifestly antisymmetric under the permutation $1\leftrightarrow 2$, which validates our prescription (\ref{wopm}) for $W_1$.

Collecting the terms (\ref{ioo}) and (\ref{ioit}), we find 
\begin{align}
 Z_3= \sqrt{C(p_1,p_2,p_3)}\ (z_1-z_2)^{\Delta_3-\Delta_1-\Delta_2} (z_2-z_3)^{\Delta_1-\Delta_2-\Delta_3} (z_3-z_1)^{\Delta_2-\Delta_3-\Delta_1}\ ,
\label{zc}
\end{align}
where we recall the relation $\Delta=p^2$ (\ref{da}) between conformal dimensions and momentums, 
and the three-point structure constant is 
\begin{align}
 C(p_1,p_2,p_3) = C_0\frac{\Upsilon(p_1+p_2+p_3)\Upsilon(-p_1+p_2+p_3)\Upsilon(p_1-p_2+p_3)\Upsilon(p_1+p_2-p_3)}{\Upsilon(2p_1)\Upsilon(2p_2)\Upsilon(2p_3)}\ ,
\label{cco}
\end{align}
using the function $\Upsilon$ which we define from Barnes' $G$-function by 
\begin{align}
 \Upsilon(x) =  e^{x^2}G(1+x) G(1-x)\ ,
\label{lug}
\end{align}
so that we have
\begin{align}
 \pp{x} \log \Upsilon(x) = x\left(\psi(x)+\psi(-x)\right)\ . 
\label{ptu}
\end{align}
Recovering $Z_3$ from the mere knowledge of $\left(\pp{p_1}-\pp{p_2}\right)\log Z_3$ is possible thanks to permutation symmetry and the reflection property (\ref{vrv}), as explained in \cite{cer12}. From \cite{cer12} we moreover expect that the resulting $Z_3$ is actually not the three-point function, but its holomorphic square root, hence the square root in eq. (\ref{zc}). Our formalism deals only with quantities which are locally holomorphic as functions of $z_i$. Holomorphicity is essential for the derivations of Ward identities, isomonodromy properties, and Seiberg-Witten equations. This restriction to holomorphic quantities is no obstacle to solving the theory, because correlation functions have decompositions (such as (\ref{zfi})) into such quantities. 

Finally, let us discuss the remaining ambiguities in $C(p_1,p_2,p_3)$, which are encoded in the factor $C_0$ in eq. (\ref{cco}). As a function of $p_i\in\R$, the factor $C_0$ should be locally constant, but not necessarily globally constant: $C_0$ can jump whenever the $\Upsilon$ factors are singular. The singularities of $\Upsilon(x)$ are apparent in the equation
\begin{align}
 \frac{\Upsilon(x+1)}{\Upsilon(x)} = \frac{\Gamma(x+1)}{\Gamma(-x)} e^{2x+1}\ ,
\label{upu}
\end{align}
which follows from $\frac{G(x+1)}{G(x)}=\Gamma(x)$, and allows us to determine $\Upsilon(x)$ globally, starting from the interval $x\in (0,1)$ where $\Upsilon(x)$ is smooth. 
The null hypothesis would now be to set $C_0=1$, thereby ensuring that $C(p_1,p_2,p_3)$ is meromorphic. But this hypothesis is not particularly well-motivated. Only the values $p\in\R-\frac12\Z$ correspond to unitary representations of the Virasoro algebra, and indeed the spectrum of Runkel-Watts theory is made of those values \cite{rw01}. (The exclusion of the half-integer values of $p$, which form a set of measure zero, does not affect the correlation functions.) Rather than requiring $C(p_1,p_2,p_3)$ to be meromorphic, we should require it to be positive when $p_1,p_2,p_3\in \R$, which is necessary for $\log Z_3$ to be well-defined. So we will allow deviations from $C_0=1$ whenever the positivity of $C(p_1,p_2,p_3)$ requires them. The factors $\Upsilon(2p_i)$ in the numerator of eq. (\ref{cco}) actually do not require any such deviations, as their positivity can be ensured by renormalizing the fields $V_{p_i}(z_i,\bar{z}_i)$. Using eq. (\ref{upu}), as well as $\Gamma(x\in\R)\in\R$ and $\Gamma(x)\Gamma(1-x)
=\frac{\pi}{\sin\pi x}$, we have
\begin{align}
 x\in\R \quad \Rightarrow \quad \mathrm{sign}\frac{\Upsilon(x+1)}{\Upsilon(x)} = \mathrm{sign}\sin\pi x\ .
\end{align}
So the contribution of the denominator of eq. (\ref{cco}) to the sign of $C(p_1,p_2,p_3)$ is 
\begin{align}
 \sigma = \mathrm{sign}\Big(\sin\pi(p_1+p_2+p_3)\sin\pi(-p_1+p_2+p_3)\sin\pi(p_1-p_2+p_3)\sin\pi(p_1+p_2-p_3)\Big)\, .
\label{ss}
\end{align}
It follows that $C_0$ must be a function of $\sigma$. By a choice of overall normalization, we can assume $C_0(+)=1$. All we can say about $C_0(-)$ is $C_0(-)\leq 0$. Our methods therefore determine $C(p_1,p_2,p_3)$ up to the choice of $C_0(-)$.

\subsection{Comparison with known results}

The three-point function of Runkel-Watts theory is most clearly written in Section 1 of \cite{sch03}, where Runkel-Watts theory is called the ``Euclidean theory''. The three-point function has a locally constant factor, which is made of step functions and which corresponds in our notations to 
\begin{align}
 C_0 = \frac{1+\sigma}{2}\ ,
\end{align}
where $\sigma$ was defined in eq. (\ref{ss}). In other words, $C_0(-)=0$. This vanishing of the three-point structure constant for half of the values of the momentums is a consequence of the fusion rules of Minimal Models in the limit $c\rightarrow 1$ \cite{rw01}. It would be interesting to check whether $C_0(-)=0$ is the only value of $C_0(-)$ which leads to a crossing-symmetric three-point function. 

The remaining factors \cite{sch03} coincide with our three-point structure constant (\ref{cco}) provided our special function $\Upsilon$ (\ref{ptu}) has an integral representation of the type
\begin{align}
 \log\Upsilon(x) = \lambda_0 + \lambda_1 x +\lambda_2 x^2- \int_0^\infty \frac{dt}{t}\left(\frac{\sinh^2 xt}{\sinh^2 t}-e^{-2t}x^2\right) \ ,
\label{lui}
\end{align}
where $\lambda_0,\lambda_1$ and $\lambda_2$ are constants. This equation indeed holds, because its third derivative can be deduced from the identity
\begin{align}
 \psi'(x) + \frac{x-1}{2}\psi''(x) = \int_0^\infty dt \frac{t^2 }{\sinh^2 t}e^{2t(1-x)}\ .
\end{align}
Our expression for the three-point structure constant is therefore simpler than the previously known formula: the function $\log\Upsilon$, which was known through the integral representation (\ref{lui}), is now shown to be the primitive of a known function by eq. (\ref{ptu}). This reflects the fact that our derivation of the three-point function is simpler. There might actually be an even simpler derivation, according to the principle that \textit{the simplest possible derivation of a formula is no more complicated than the formula itself}.

\section{Conclusion \label{seccon}}

It had been shown in \cite{cer12} that Liouville correlation functions are encoded in certain spectral curves. In the case $c=1$, we have now found even more basic objects: Lax matrices. In contrast to some constructions in the literature, these objects come with methods for explicitly computing the correlation functions. In particular, a Lax matrix allows us to build exact solutions of the Ward identities, whereas a spectral curve encodes a perturbative expansion. It would thus be very interesting to generalize the Lax matrix formalism to arbitrary values of the central charge $c$. In other words, how do we generalize the determinantal ansatz of Section \ref{ssecmat} when $c\neq 1$, and solve the Ward identities of Section \ref{ssecwi}? This important technical question has no known answer. 

On the other hand, the construction (\ref{llt}) of the Lax matrix as the matrix square-root of the energy-momentum tensor, which encodes the Virasoro symmetry algebra, can surely be generalized to higher symmetry algebras, such as W-algebras. This might help solve the problem of computing correlation functions in conformal Toda theory -- for a particular value of the central charge. Solving conformal field theories is vastly more complicated when higher symmetry algebras are involved: only with the Virasoro algebra do all correlation functions follow from correlation functions of primary fields. The correlation functions $W_m$ of Section \ref{ssecwi}, which we used as auxiliary quantities, might be very helpful in encoding correlation functions of descendent fields.



\acknowledgments{
We are grateful to Michel Berg\`ere, Leonid Chekhov, Robert Conte and Oleg Lisovyy for useful discussions and debates. We also thank the anonymous JHEP referee for his work.
}


\end{document}